\documentclass[letter]{aa} 

\usepackage{graphicx}
\usepackage{txfonts}
\def\HI {H\kern0.1em{\sc i}}
\def\hii {H\kern0.1em{\sc ii}}

\def\kms{km s$^{-1}$}

\def\solum {\hbox{L$_{\odot}$}}

\usepackage{natbib}

\begin{document}

   \title{Detection of maser emission at 183 and 380 GHz with ALMA in the gigamaser galaxy TXS\,2226-184}


   \author{A. Tarchi\inst{1}
          \and
          P. Castangia\inst{1}
          \and
           G. Surcis\inst{1}
           \and
           V. Impellizzeri\inst{2}
           \and
           E. Ladu\inst{1,3}
           \and
           E. Yu Bannikova\inst{4,5,6}
          }

   \institute{INAF-Osservatorio Astronomico di Cagliari, via della Scienza 5, 09047, Selargius (CA), Italy
             \email{andrea.tarchi@inaf.it}
          \and
          Leiden Observatory, Leiden University, Post Office Box 9513, 2300 RA Leiden, Netherlands            
          \and
Department of Physics, University of Cagliari, S.P.Monserrato-Sestu km 0,700, I-09042 Monserrato (CA), Italy
          \and
INAF - Astronomical Observatory of Capodimonte, Salita Moiariello 16, Naples I-80131, Italy
          \and
Institute of Radio Astronomy, National Academy of Sciences of Ukraine, Mystetstv 4, Kharkiv UA-61002, Ukraine
          \and
V.N.Karazin Kharkiv National University, Svobody Sq.4, Kharkiv UA-61022, Ukraine}

   \date{Received gg mm yyyy; accepted gg mm yyyy}

 
  \abstract
   {The low-ionization nuclear emission-line region (LINER) galaxy TXS\,2226-184 is known to host a very luminous 22 GHz water maser, called a  gigamaser at the time of its discovery. To date, the nature of this maser is still being debated, in particular, whether it is associated with a nuclear accretion disk or with an ejection component, namely a jet or an outflow originating in the active galactic nucleus.}
   {We   obtained multi-band (bands 5, 6, and 7) ALMA observations during Cycle 9, with the purpose of investigating the maser nature and the nuclear molecular material in the innermost region of the galaxy.}
   {While the full data sets are still under study, a preliminary data reduction and analysis of the band 5 and 7 spectral line cubes presented in this Letter  already offer a significant outcome.} 
   {We observed bright, possibly maser emission from the water 183 GHz and 380 GHz transitions in TXS\,2226-184. To the best of our knowledge, this represents the first unambiguous detection (S/N $\ge$ 100) of 380 GHz maser emission in a known 22-GHz maser galaxy, and the first case where all  three transitions are present in the same object. Emission features at both frequencies show a two-peaked line profile resembling that of the 22 GHz maser features. The millimeter/submillimeter emission originates from a region coincident, within the errors, with that of the 22 GHz.}
   {The similarities in profile and position indicate that the emission at the three frequencies is likely produced by the same nuclear structure, although differences in line strengths and feature peak positions may hint at a slightly different physical conditions of the emitting gas. A comparison with the few megamaser sources studied at high enough detail  and sharing similarities with the water lines in TXS\,2226-184 favors a nature associated with the amplification of a bright nuclear continuum (from a jet or outflow) through dense and hot gas in front of the nucleus (e.g., a disk or torus); however, a more comprehensive analysis of the available data is necessary to better assess this scenario.}


\titlerunning{ALMA observations of TXS2226-186}
\authorrunning{Tarchi A. et al.}

   \maketitle


\section{Introduction}

Water masers associated with active galactic nuclei (AGN) and known as megamasers have been related with
three distinct phenomena: (i) with nuclear accretion disks, where they can be used to derive the disk
geometry, enclosed nuclear mass, and distance to the host galaxy (see, e.g., \citealt{Miyoshi1995} and \citealt{Herrnstein1999} for NGC\,4258, and, more recently, \citealt{Braatz2010} for UGC\,3789), or
sometimes with the innermost boundary of a dusty torus in the region of the interaction with the outflows
(e.g., \citealt{Bannikova2023}); (ii) with radio jets, where they can provide important information about the evolution of jets
and their hotspots (e.g., \citealt{Peck2003} for Mrk 348, and, more recently, \citealt{Castangia2019}, for IRAS 15480); (iii) with nuclear
outflows, tracing the velocity and geometry of nuclear winds at < 1 pc from the nucleus, as in the
case of Circinus (\citealt{Greenhill2003}) and NGC 3079 (\citealt{Kondratko2005}), where they offer a promising means to probe the structure
and motion of the clouds in the toroidal obscuring region predicted by clumpy torus models
(e.g., \citealt{Nenkova2008}), and to help studies of AGN tori in general. In all these cases, water megamasers provide vital information on the structures, dynamics
and composition of the molecular gas at close proximity to the AGN.
In addition, thermal lines can provide complementary information to the megamasers.
Recent Atacama Large Millimeter/submillimeter Array (ALMA) observations of the molecular interstellar medium (ISM), and, in particular, of the CO(6-5), HCN(J=3-2), and HCO$^{+}$(J=3-2) transitions, in the central few pc of nearby AGN have revealed complex kinematics (\citealt{GarciaBurillo2016}; \citealt{Gallimore2016}; \citealt{Imanishi2018}; \citealt{Impellizzeri2019}).

Most previous observational work on water vapor megamasers has focused on the $6_{1, 6}-5_{2, 3}$ transition at a rest frequency of 22.235079 GHz\footnote{Transitions and frequencies are from Splatalogue: https://splatalogue.online/\#/home} (hereafter referred to as the 22 GHz line). Thanks to this transition, the relation with the three aforementioned phenomena has already been studied.

However, radiative transfer models and observations have found that similar conditions also give rise to additional water maser lines in the millimeter/submillimeter band. Megamaser detections have been obtained in a growing number of galaxies in 183, 321, and 325 GHz water maser transitions (e.g., \citealt{Humphreys2016}; \citealt{Pesce2023}). These studies suggest that (sub)millimeter water masers may be common in AGN, associated with nuclear activity like the 22 GHz megamasers sources. Because of the matching velocities, the masers seem to occur in broadly the same regions as the 22 GHz, originating in the same thin, edge-on disk delineated by the 22 GHz lines, with  submillimeter emission lines detected both at systemic and high rotation velocities. The line ratios of two or more maser transitions originating from the same gas would constrain radiative transfer models far better than is now possible (e.g., \citealt{Humphreys2005}). The $3_{1, 3}-2_{2, 0}$ transition at a rest frequency of 183.310087  (hereafter the 183 GHz line)  and the $4_{1, 4}-3_{2, 1}$ transition at a rest frequency of 380.1973598 GHz (hereafter the 380 GHz line), particularly together with that at 22 GHz, are known as the ``backbone'' masers, and are among the most promising water maser lines predicted by,  \citet{Neufeld1991} and \citealt{Gray2016}, among others.

TXS\,2226-184 (hereafter TXS\,2226) is located at a distance of 107 Mpc ($z$ = 0.025; \citealt{Kuo2018}), is optically classified as an elliptical/S0 galaxy, and
spectroscopically identified as a low-ionization nuclear emission-line region (LINER), which is a signature of a relatively low-luminosity active nucleus. In this galaxy, a bright 22 GHz water maser was detected in 1995 by \citet{Koekemoer1995}, and the reported emission was so luminous (L$_{iso}$
 = 6100 \solum), given its redshift, that the term ``gigamaser” was coined. The detection of such a
bright maser source in an early-type galaxy was quite unexpected. The bright nuclear
radio emission, although heavily obscured, shows a symmetric and tight structure, seemingly produced by jets oriented at a large angle to the line of sight \citep{Taylor2004}. 
A very recent detailed very long baseline interferometry (VLBI) study, performed with the Very Long Baseline Array (VLBA) and the European VLBI network (EVN), of the 22 GHz water gigamaser in the nucleus of
TXS 2226-184 has associated, for the first time, the water maser features with the most luminous
radio continuum clump reported in the literature, and a speculative model for the maser origin has
also been proposed \citep{Surcis2020}.

During Cycle 9, we observed the nuclear region of TXS\,2226 with ALMA at Bands 5, 6, and 7. In the following, we present the first results from the band 5 and 7 measurements, and discuss their impact in the framework of our understanding of the  complex  scenario present in the nucleus of this gigamaser galaxy.

\section{Observations and data reduction}
\label{sect:obs}

The observations whose outcome is reported in this paper were performed as part of the ALMA project 2022.1.01591.S (PI: Tarchi).
ALMA was used to observe the innermost region of TXS\,2226 with dual polarization at Band 5 on May 24, and Band 7 on June 9, 2023. On April 28, 2023, Band 6 was also observed, but it is not included in this work.
Observations of the target were interleaved with observations of calibrators used to determine
bandpass, flux density, and station gain calibration. 

The spectral setup for each band consisted of four spectral windows. Spectral windows were configured to have a frequency resolution of $\sim$ 1.13 MHz and a width in frequency of $\sim$ 2 GHz (corresponding to 1.6 and 0.8 \kms\, and 3000 and 1500 \kms\ for band 5 and 7, respectively). In particular, one spectral window was centered around the expected locations of the maser lines.

All data from this ALMA project have a corresponding quality assurance 2 (QA2) data processing. We   carried out the imaging of the spectral line continuum-subtracted data sets using the task `tclean' of the Common Astronomy Software Applications (CASA) software\footnote{https://casa.nrao.edu/}, version 6.5.4.9. The restoring beam size of the images is 0.13 and 0.06 arcsec, for the 183 and 380 GHZ cubes, respectively. We extracted the spectra from the image cubes using the spectral profile tool. In addition, we retrieved a data set from the Green Bank Telescope (GBT) archive. This was observed on December 9, 2010, under the project name AGBT10C\_013. The details of this observation are thoroughly reported in \citet{Surcis2020}, and will not be replicated here. The spectra obtained were subsequently imported in the Continuum and Line Analysis Single-dish Software (CLASS) package, part of the Grenoble Image and Line Data Analysis Software (GILDAS)\footnote{https://www.iram.fr/IRAMFR/GILDAS} for follow-up investigation (e.g., Gaussian fitting, comparison).


\section{Results}

\begin{figure}
\includegraphics[width=8.9cm]{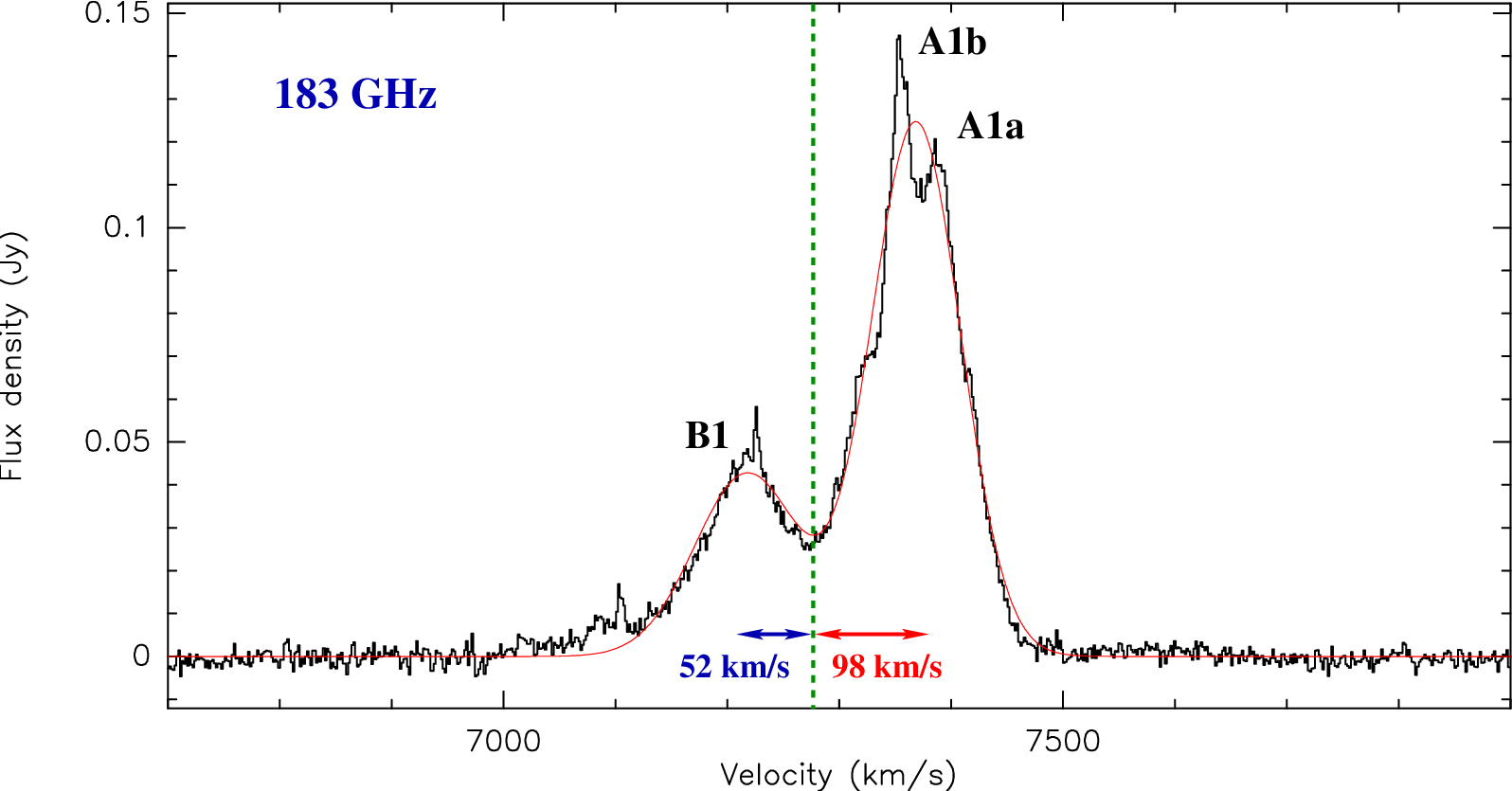}
\caption{Spectrum  of the 183 GHz maser in TXS\,2226. The dashed green line indicates the systemic velocity of the galaxy in the radio convention (7270 \kms). The red and blue arrows indicate the reported offsets (in \kms) of the feature's peak velocity (taken from the Gaussian fit in red here) from the systemic velocity.}
\label{fig:maser183}
\end{figure}

The outcome of our data analysis yielded two important results. We detected spectral line emission at 183 GHz in TXS\,2226. Figure~\ref{fig:maser183} shows the spectra obtained from the ALMA cube. In particular, the emission has a clear double-peak structure with very broad profile. The line profile displays two main spectral features (labeled A1 and B1) with a peak flux density of 125 and 43 mJy, and a width of 94 and 110 \kms, respectively.

By using the standard equation (see, e.g., \citealt{Humphreys2016}; their Eq. 1) to derive the isotropic luminosity of the (maser) emission, L$_{iso}$, we derived values of $\sim$ 25000 and 10000 \solum\ for A1 and B1, respectively. These values fall in ranges typical for sources classified as gigamasers. The results for a Gaussian fit of the individual spectral line features are reported in Table~\ref{tab:maserfits}.



\begin{figure}
\includegraphics[width=8.9cm]{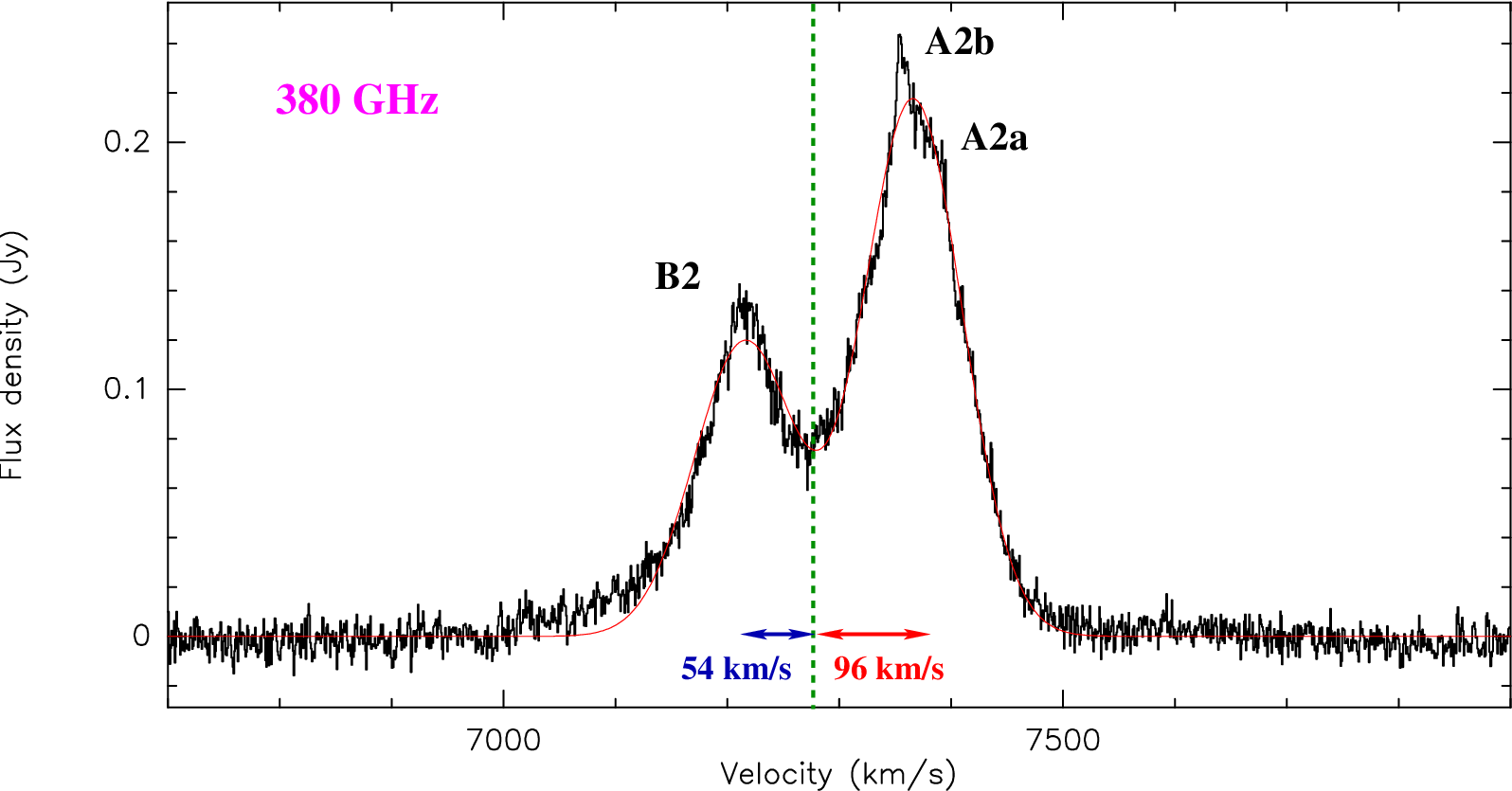}
\caption{Spectrum  of the 380 GHz maser in TXS\,2226. The dashed green line indicates the systemic velocity of the galaxy in the radio convention (7270 \kms). The red and blue arrows indicate the reported offsets (in \kms) of the feature's peak velocity (taken from the Gaussian fit in red here) from the systemic velocity.}
\label{fig:maser380}
\end{figure}


We also detected, for the first time, spectral line emission at 380 GHz in TXS\,2226 (Fig.~\ref{fig:maser380}). Also in this case, the emission has a double-peak structure with a very large width. The two features displayed at 380 GHz (labeled A2 and B2) have similar widths (both of order 100 \kms) to those at 183 GHz, but higher peak flux densities (217 and 119 mJy, respectively). Their isotropic luminosities are $\sim$ 100000 and 60000 \solum, for A2 and B2, respectively. These values indicate extremely bright sources, seemingly classifiable as gigamasers. Fit values for the individual features are reported in Table~\ref{tab:maserfits}.

\begin{table*}
\begin{center}
\caption{Parameters of the maser features detected in TXS\,2226 obtained via the Gaussian fit performed in CLASS.}  
\label{tab:maserfits}
\centering
\begin{tabular}{lccccc}
\hline \hline
  Feature  &  Peak flux density     &       FWHM         &      Area                     & Peak velocity  & L$_{iso}$ \\
          & (Jy)   &  (\kms)  & (Jy $\cdot$ \kms) & (\kms) & \solum \\
\hline
A1$(a+b)$ & 0.125 & 94.2 (  0.2) &  12.48     (  0.03) & 7368.7 (  0.2)  & 25589 \\
B1 & 0.043 & 110.2 (  0.8) &  5.01     (  0.03) & 7217.9 (  0.4)  & 10281 \\
\hline
A2(a$+$b) & 0.217 & 103.6 (  0.5) &  24.0     (  0.1) & 7366.2 (  0.2) & 101893 \\
B2 & 0.119 & 106 (  1) &  13.5     (  0.1) &  7215.8 (  0.4) &  57465 \\
\hline
\end{tabular}
\tablefoot{The columns indicate: the spectral line (maser) feature label; the peak flux density; the feature full width at half maximum, area, and peak velocity; the isotropic luminosity of the feature.}
\end{center}
\end{table*}

\section{Discussion}

The actual angular resolution of our observations at 183 GHz is 0.13 arcsec. The lower limit to the brightness temperature of our spectral line emission is then of order 130 K, and thus it is impossible to determine on a T$_{b}$ basis whether the emission is thermal or nonthermal (maser). However, for the emission to be thermal, but with a brightness temperature  $\le$ 2000 K, such that the water molecules are not dissociated, the emitting region has to be $>$ 0.03 arcsec (corresponding to 15 pc), which would be significantly more extended than expected from previous similar cases (see, e.g., \citealt{Humphreys2005}).

Even with the resolution of our 380 GHz measurements (0.06 arcsec), the lower limit to the brightness temperature of our spectral line emission ($\sim$ 400 K) is still too low to rule out a thermal nature of the emission. In this case, the emitting region would have to be $>$ 0.05 arcsec (corresponding to 25 pc), once again not as compact as possibly expected.

In addition, by considering the similar profiles of all emitting lines (see next section), and that the 22 GHz emission in TXS\,2226 has been confidently associated with the maser phenomenon \citep{Surcis2020}, it is natural to assume that  the 183 and 380 GHz features also have a maser nature. In addition, the presence of strong 22 GHz, and relatively strong 183 and 380 GHz emission indicate conditions for the gas of relatively high density ($n_{H}$ $>$ $10^{4-6}$ cm$^{-3}$) and/or high temperature ($>$ 50-100 K), different from those found, for example, in Arp220 \citep{Cernicharo2006}. This fact and the high luminosity of the  maser  features seemingly rule out a star formation origin (see also, \citealt{hagiwara2013}), thus confirming a nature for the detected features associated with AGN activity in the target.

In the following, we     discuss the spectral line features according to the aforementioned scenario, although we are aware that a different one cannot be ruled out a priori. 
 
\subsection{Comparison of the (maser) features}

\begin{figure*}
\centering
\includegraphics[width = 16 cm]{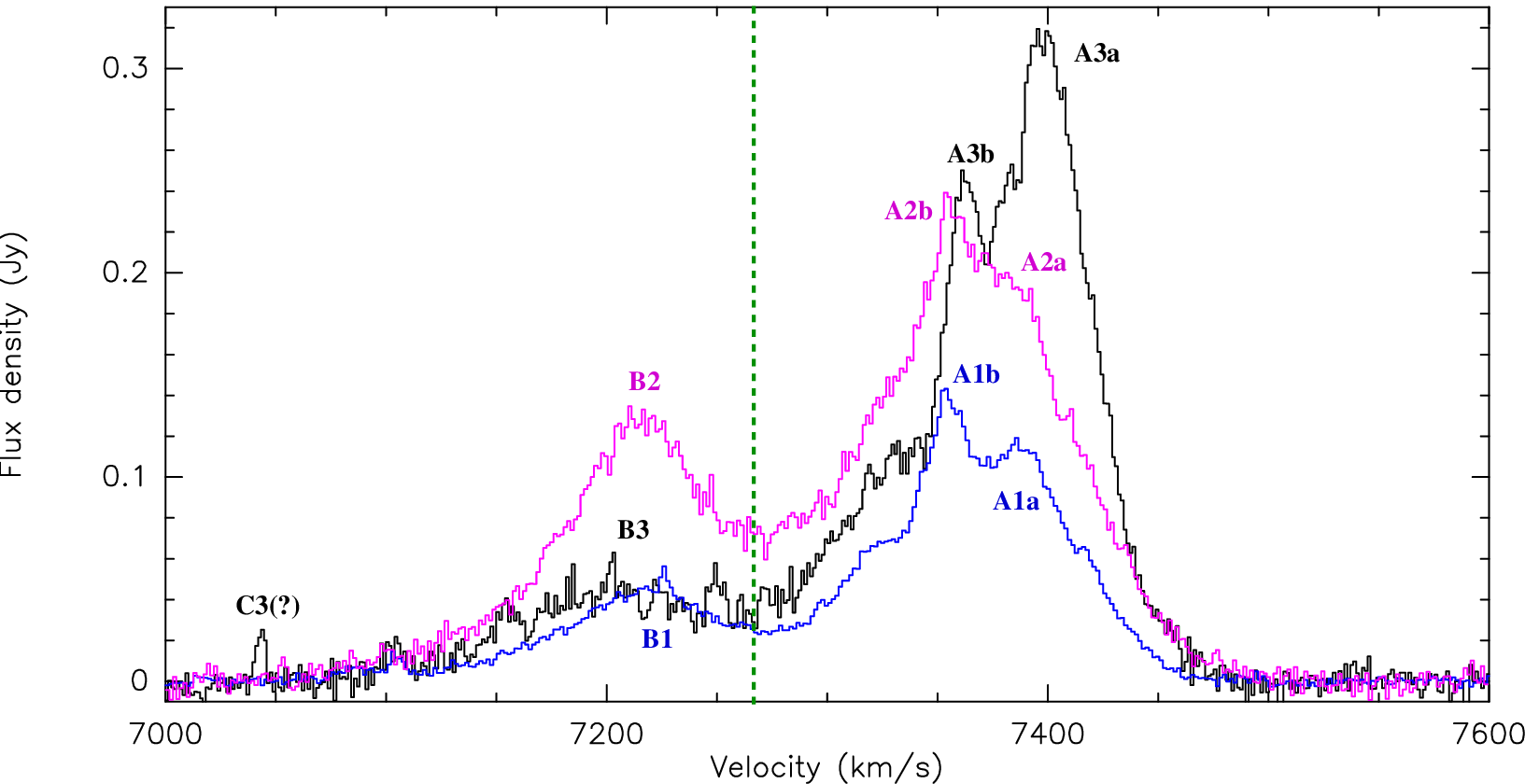}
\caption{Comparison of the spectra of the maser emission at 22 (black), 183 (blue), and 380 (magenta) GHz in TXS\,2226, after smoothing the spectra to the coarsest spectral resolution of the 183 GHz data ($\Delta$v $\sim$ 1.6 \kms). The dashed green line indicates the systemic velocity of the galaxy in the radio convention (7270 \kms).}
\label{fig:comparison}
\end{figure*}

Figure~\ref{fig:comparison} shows a plot where all maser emission detected at the three different frequencies, 22 (with archival GBT measurements; see Sect.~\ref{sect:obs}), 183, and 380 GHz (with ALMA, present work), are displayed.
Given that megamaser sources may be variable and the spectra are taken at different epochs (in particular, more than ten years have elapsed since the GBT observations), we note that  only a qualitative comparison is performed here. However, together with the unique chance of having three (maser) transitions observed in one target, the relative stability 
 of the line profile and intensity of the 22 GHz maser emission in TXS\,2226 over a 20-year period, indicated by multi-epoch (though not frequent and/or regular) GBT spectra (e.g., \citealt{Surcis2020}, and references therein), make such a comparison potentially relevant.     
The broad velocity range covered by the emission is consistent between the different transitions, indicating that they arise from the same region. Some differences are visible, however. 
In particular, the flux densities are different between the 22 GHz and 183 GHz line:  the former is more intense for feature A by almost a factor of three, although comparable in strength for feature B. 
A 183 GHz transition weaker then the 22 GHz transition is expected since this is what has also been found in other extragalactic sources (e.g., \citealt{Pesce2023}), despite  differences of up to an order of magnitude. 
The strength of the 380 GHz transition is instead more difficult to interpret since to the best of our knowledge this is the first confident detection (S/N $\ge$ 100) of 380 GHz maser emission in a known 22 GHz maser galaxy, and the first case where all three 
 transitions are found. The 380 GHz line falls outside the ALMA band 7 frequency range, and it is detected in TXS\,2226 thanks to the cosmological redshift that relocates it to within the observable range. A first detection of this maser was obtained by \cite{Phillips1980} with the {\it Kuiper} Airborne Observatory in the Becklin-Neugebauer Kleinmann–Low Nebula of Orion. A tentative detection of a broad (possibly maser) line feature was also reported by \cite{Kuo2019} and re-analyzed by \cite{Stacey2020} in the high-z lensed QSO MG\,J0414+0534. In addition, 380 GHz emission has been reported in two lensed galaxies, APM\,08279+5255 and NCv1.143 (S/N of $\sim$ 11 and 6, respectively), where however, no significant 22 GHz maser emission was observed \citep{Yang2023}.
The 380 GHz line belongs, together with the 22 and 183 GHz lines, to the maser transitions involving the backbone levels. The strength of the 380 GHz line emission is then expected to be large and, in principle, depending on the degree of maser saturation, even brighter than the 22 GHz emission \citep{Gray2016}. This seems to be the case for feature B, while for feature A the 380 GHz flux density is still below that at 22 GHz. The 380 GHz emission is, however, noticeably brighter than the 183 GHz line.
The majority of the features detected at 22 GHz are also generally present at the other two frequencies (Fig.~\ref{fig:comparison}). In particular, the presence of the feature labeled B3 (B1 and B2 at 183 and 380 GHz, respectively), and of the feature A3 (namely, A1 and A2) have been confirmed. Interestingly, the 22 GHz feature A3a seems to be absent or greatly reduced in flux density in the ALMA spectra. Overall the millimeter/submillimeter maser main features seems to peak at a velocity slightly closer to the systemic velocity of the galaxy than the 22 GHz counterparts (Fig.~\ref{fig:comparison}). Differences in peak velocity of different maser transitions may be due to diverse pump mechanisms and/or geometry associated with each maser source (see, e.g., \citealt{Humphreys2005}). Also a slightly different location of the higher-frequency masers with respect to those at  22 GHz, for example  at smaller or larger radius or distance from the putative center of activity, with different gas density or temperature conditions would also lead to this spectral profile comparative scenario (see, e.g., \citealt{Pesce2023}, for the Circinus case). In addition, given the gap of more than ten years between the observations when the 22 GHz and ALMA spectra were taken, another potential explanation could be the motion of the emitting molecular gas during the intervening time. Unfortunately, the spatial resolution is too coarse to yield detailed position estimates of the different maser emitting regions. 

\subsection{The possible nature of the maser emission}
\label{sect:nature}

In \citet{Surcis2020} the possible nature of the 22 GHz maser is investigated through VLBI observations. Their conclusion  clearly indicated a nuclear origin of the emission, associated with the AGN activity of TXS2222. However, it also left open the possibility that the maser was associated with either a jet--outflow component or, although with reduced confidence, to an accretion disk.

\begin{figure*}
\centering
\includegraphics[width = 16 cm]{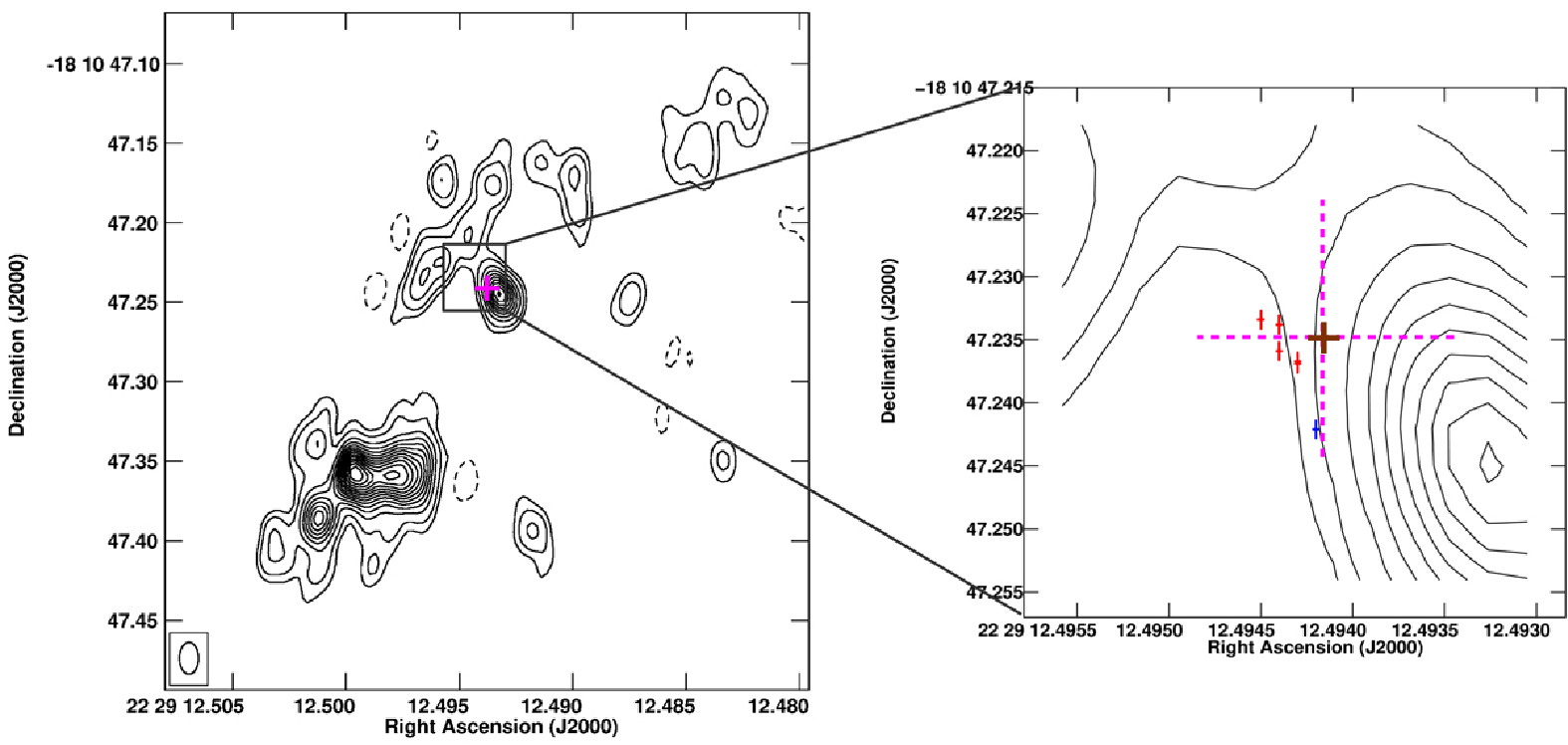}
\caption{Adaption of  Fig. 9  in \citet[]{Surcis2020} with the (coincident) positions of the maser emission at 183 and 380 GHz (magenta cross) overlaid on the VLBI contour map of the L-band continuum emission in TXS\,2226 \citep{Taylor2004}. The size of the magenta cross (solid and dashed, in the left and right panel, respectively) is comparable with the absolute astrometric precision estimated for our ALMA observations ($\sim$ 10 mas; see Sect.~\ref{sect:nature}). The size of the brown cross in the right panel is the relative position uncertainty of the (sub)millimeter masers (1 mas; Sect.~\ref{sect:nature}). The red and blue crosses are the 22 GHz maser spots detected through VLBA measurements (for details, see \citealt{Surcis2020}, and references therein).}
\label{fig:location}
\end{figure*}

More recently, \cite{Tarchi2023} reported the preliminary results of a multi-band radio continuum campaign led with the EVN that indicates that the 22 GHz maser spots seem to be associated with a region for which the radio spectrum is inverted and opacity is high, thus pinpointing the nucleus of the galaxy. The assessment of the nature of the maser requires a more thorough analysis of these VLBI observations and their combination with the present ALMA results that is beyond the purpose of this letter and that, together with a theoretical model of the AGN components, will be included in a subsequent publication (Tarchi et al. in preparation). Some details of the present millimeter/submillimeter detections may, however, already anticipate interesting clues to delineate the maser origin. 

First of all, the position derived from centroid measurements of both the 183 and 380 GHz masers is coincident, within the statistical uncertainty of 1 mas, at RA$_{J2000}$=$\rm{22^{h}:29^{m}:12.4942^{s}}$ and Dec.$_{J2000}$=$\rm{-18^{\circ}:10':47.235''}$. Considering that the absolute astrometric precision for our ALMA observations is on the order of 10 mas (see, e.g., \citealt{Pesce2016}, and Sect.\,A.9.5 of the ALMA Cycle\,9 Proposer’s Guide), the location of the (sub)millimeter masers appears to be coincident with that of the 22 GHz maser spots derived by Surcis, Tarchi \& Castangia (2020; Fig.~\ref{fig:location}). This co-spatiality of the emitting regions and the similar overall spectral structure (see previous section) suggest that all features are tracing roughly the same gas and support the maser nature of the emission.

By comparing the case of TXS\,2226 with other megamaser sources, the most striking characteristic that stands out is the variety of widths. In particular, for masers associated with AGN activity, usually those produced in accretion disks show relatively narrow features bundled in three groups: at velocities close to the recessional velocity of the galaxy, and redshifted  and blueshifted with respect to this velocity. For masers related to the AGN ejection activity (e.g., jet and/or outflows), the situation is less ordered, but often in these cases broad (usually) redshifted feature are observed (see, e.g., \citealt{Tarchi2012}). While an unambiguous interpretation for the maser nature based only on the maser profiles is not possible (composite and peculiar maser profiles are, in fact, not rare), and  a detailed map of the maser spot distribution is usually necessary, it is   reasonable to compare and discuss the cases that show the greatest similarities between the millimeter and submillimeter maser profiles. To date, only one source showing 22 GHz broad maser lines similar to those displayed by TXS\,2226, and lacking high-velocity components has been reported to host maser emission also at millimeter wavelengths (although at 321 GHz): the radio galaxy NGC\,1052 \citep{Kameno2023b}. For the 321 GHz spectrum, the authors report a two-peaked profile emission line that is, however, quite different (much weaker and displaced in velocity) from that at 22 GHz. Nevertheless, their conclusions hint at a nature for the 321 GHz maser emission associated with the amplification of a bright nuclear continuum (from a jet or outflow) through dense and hot gas in front of the nucleus (e.g., a disk or torus). The presence of a relatively strong nuclear radio jet  in TXS\,2226 makes a similar scenario reasonable, even for the millimeter maser emission in this galaxy. This would also be consistent with the interpretation presently invoked also for the 22 GHz maser in TXS\,2226 by \citet{Surcis2020} and \cite{Tarchi2023}. However, a more detailed study of the nuclear components, both at centimeter and millimeter wavelengths, in TXS\,2226 is indeed necessary to draw a conclusion.

\begin{acknowledgements}
We thank the anonymous referee for his/her insightful suggestions. The authors would like to acknowledge the European ARC, and, in particular, Kazi Rygl for their help during several phases of the project. AT is grateful to his daughter Elena for her constant encouragement.
This paper makes use of the following ALMA data: ADS/JAO.ALMA\#2022.1.01591.S. ALMA is a partnership of ESO (representing its member states), NSF (USA) and NINS (Japan), together with NRC (Canada), MOST and ASIAA (Taiwan), and KASI (Republic of Korea), in cooperation with the Republic of Chile. The Joint ALMA Observatory is operated by ESO, AUI/NRAO and NAOJ.
\end{acknowledgements}

\bibliographystyle{aa} 
\bibliography{Bibliografia_Tarchi_new} 


\end{document}